# Last giant impact on the Neptunian system

## Constraints on oligarchic masses in the trans-Saturnian region

M. Gabriela Parisi[1,2,*] and Luciano del Valle[3,**]

[1] Instituto Argentino de Radioastronomía (IAR-CONICET), C.C. N$^o$ 5, 1894 Villa Elisa, Argentina
    e-mail: `gparisi@iar-conicet.gov.ar`
[2] Facultad de Ciencias Astronómicas y Geofísicas, Universidad Nacional de La Plata, Argentina
[3] Departamento de Astronomía, Universidad de Chile, Casilla 36-D, Santiago, Chile
    e-mail: `ldelvall@das.uchile.cl`

**Abstract.**

*Context.* Current models of the formation of ice giants attempt to account for the formation of Uranus and Neptune within the protoplanetary disk lifetime. Many of these models calculate the formation of Uranus and Neptune in a disk that may be several times the minimum mass solar nebula model (MMSN). Modern core accretion theories assume the formation of the ice giants either in situ, or between ∼ 10-20 AU in the framework of the Nice model. However, at present, none of these models account for the spin properties of the ice giants.

*Aims.* Stochastic impacts by large bodies are, at present, the usually accepted mechanisms able to account for the obliquity of the ice giants. We attempt to set constraints on giant impacts as the cause of Neptune's current obliquity in the framework of modern theories. We also use the present orbital properties of the Neptunian irregular satellites (with the exception of Triton) to set constraints on the scenario of giant impacts at the end of Neptune formation.

*Methods.* Since stochastic collisions among embryos are assumed to occur beyond oligarchy, we model the angular momentum transfer to proto-Neptune and the impulse transfer to its irregular satellites by the last stochastic collision (GC) between the protoplanet and an oligarchic mass at the end of Neptune's formation. We assume a minimum oligarchic mass $m_i$ of 1 $m_\oplus$.

*Results.* From angular momentum considerations, we obtain that an oligarchic mass $m_i \sim 1 m_\oplus \leq m_i \leq 4\ m_\oplus$ would be required at the GC to reproduce the present rotational properties of Neptune. An impact with $m_i > 4\ m_\oplus$ is not possible, unless the impact parameter of the collision were very small. This result is invariant either Neptune had formed in situ or between 10-20 AU and does not depend on the occurrence of the GC after or during the possible migration of the planet. From impulse considerations, we find that an oligarchic mass $m_i \sim 1 m_\oplus \leq m_i \leq 1.4\ m_\oplus$ at the GC is required to keep or capture the present population of irregular satellites. If $m_i$ had been higher, the present Neptunian irregular satellites had to be formed or captured after the end of stochastic impacts.

*Conclusions* The upper bounds on the oligarchic masses (4 $m_\oplus$ from the obliquity of Neptune and 1.4 $m_\oplus$ from the Neptunian irregular satellites) are independent of unknown parameters, such as the mass and distribution of the planetesimals, the location at which Uranus and Neptune were formed, the Solar Nebula initial surface mass density, and the growth regime. If stochastic impacts had occurred, these results should be understood as upper constraints on the oligarchic masses in the trans-Saturnian region at the end of ice planet formation and may be used to set constraints on planetary formation scenarios.

## 1. Introduction

The origin of the rotational properties of the planets in the Solar System is one of the fundamental questions in cosmogony. When planetesimals are accreted, they deliver rotational angular momentum caused by their relative motion with respect to the protoplanet. The angular momentum, **L**, acquired in an individual collision can be in any direction. However, because of the symmetry of the system about the plane of the planet's orbit, $(z, \dot{z}) \rightarrow (-z, -\dot{z})$, there is no systematic preference of positive or negative $L_x$ or $L_y$. An ordered component to $L_z$ is possible, thereby producing a net planetary spin angular momentum either in the same direction as the orbital angular momentum (prograde rotation) or in the opposite direction (retro-

*Send offprint requests to*: M. Gabriela Parisi
 * Member of the Carrera del Investigador Científico, Consejo Nacional de Investigaciones Científicas y Técnicas (CONICET), Argentina
 ** Fellow of the Comisión Nacional de Investigación Científica y Tecnológica (CONICYT), Chile

grade rotation). The angular momentum accreted from an uniform and dynamically cold disk of small planetesimals results in a slow systematic (ordered) component of planetary rotation in the retrograde direction (Lissauer & Kary 1991). Lissauer et al. (1997) have shown that systematic prograde rotation can be achieved if the disk density profiles are imposed such that the surface mass density near the outer edges of a protoplanet's feeding zone is significantly greater than that in the rest of the accretion zone. Schlichting & Sari (2007) obtained that planetesimals close to a meter in size, are likely to collide within the protoplanet's sphere of influence, thereby creating a prograde accretion disk around the protoplanet. The accretion of such a disk results in the formation of protoplanets spinning in the prograde sense. However, these models account for a net $L_z$ component of the planetary spin, and the problem of the obliquity (the angle between the rotational axis of a planet and the perpendicular to its orbital plane) of the ice planets remains open.

Several theories have been proposed to account for the obliquity of giant planets. A random component of planetary rotation may be due to stochastic impacts with large bodies and may be in any direction (Lissauer & Safronov 1991, Chambers 2001). Greenberg (1974) and Kubo-Oka & Nakazawa (1995) investigated the tidal evolution of satellite orbits and examined the possibility that the orbital decay of a retrograde satellite leads to the high obliquity of Uranus, but the high mass required for the hypothetical satellite makes this very implausible. An asymmetric infall or torques from nearby mass concentrations during the collapse of the molecular cloud core leading to the formation of the Solar System could twist the total angular momentum vector of the planetary system. This twist could generate the obliquities of the outer planets (Tremaine 1991). This model has the disadvantages that the outer planets must form before the infall is complete and that the conditions for the event that would produce the twist are rather strict. The model itself is difficult to be quantitatively tested. In the Nice model of Tsiganis et al. (2005), close encounters among the giant planets produce large orbital eccentricities and inclinations that were subsequently damped to the current value by gravitational interactions with planetesimals. The obliquity changes because of the change in the orbital inclinations. Since the inclinations are damped by planetesimals interactions on timescales that are much shorter than the timescales for precession due to the torques from the Sun, especially for Uranus and Neptune, the obliquity returns to low values, if it is low before the encounters (Hoi et al. 2007). Boué & Laskar (2010) reported numerical simulations showing that Uranus's axis might be tilted during the giant planet instability phase described in the Nice model, provided that the planet has an additional satellite and a temporary large inclination being the satellite ejected after the tilt. However, the required satellite is too massive. For Saturn, a good case can be made for spin-orbit resonances (Hamilton & Ward 2004), but giant impacts in the late stages of the formation of Uranus and Neptune remain the plausible explanation for the obliquities of the ice giants (Lee et al. 2007).

If giant impacts are responsible for the obliquities of the ice planets, such impacts will strongly affect the orbits of hypothetical preexisting satellites. The impulse imparted at any collision would have produced a shift in their orbital velocity. Satellites on orbits with too large a semimajor axis escape from the system (Parisi & Brunini 1997), while satellites with a smaller semimajor axis may be pushed to outer or inner orbits acquiring greater or lower eccentricities depending on the impactor mass and velocity, the initial orbital elements, the geometry of the impact and the satellite position at the moment of impact. The present physical and dynamical properties of irregular satellites may be used to set constraints on the scenario of giant impacts at the end of the formation of ice giants (Brunini et al. 2002, Maris et al. 2007, Parisi et al. 2008).

We investigate whether our results obtained for Uranus (Parisi & Brunini 1997, Brunini et al. 2002, Parisi et al. 2008) are similar when an improved method is applied to Neptune. Since stochastic collisions among embryos are assumed to occur beyond oligarchy, we model the last stochastic collision (hereafter, GC) between the protoplanet and an oligarchic mass computing the angular momentum transfer to proto-Neptune in section 2 and the impulse transfer to its irregular satellites (with the exception of Triton) in section 3. In section 4, we describe and discuss modern scenarios of planetary formation in the outer Solar System. The conclusions of the results are presented in section 5.

## 2. The spin of Neptune: Angular momentum transfer to Neptune by the last giant collision (GC)

Beyond oligarchy, the final stage of planet formation consist of close encounters, collisions, and accretion events among oligarchs. Strong impacts deliver spin angular momentum to the final planet in a random-walk fashion (Lissauer & Safronov 1991). The planetary spin accumulated by successive collisions with a distribution of small or/and large planetesimals requires ad-hoc assumptions about unknown properties of the planetesimal disk, such as the mass distribution of the bodies, the velocities distribution, and the regime of growth. We avoid the necessity of quantifying these unknown parameters modeling what happened to the planet just before it acquires its present rotational status, which is our available data.

The last off-centre giant collision (GC) between proto-Neptune and the last colliding oligarch is computed assuming that the present spin properties of Neptune are acquired by the GC. The rotational parameters of the proto-Neptune prior to the GC are random; i.e., the rotational period and the spin obliquity of Neptune before the GC are unknown and taken as initial free parameters. The procedure by Parisi et. al. (2008) is improved here, since no hypothesis is applied to the spin of Neptune prior to the GC. We also take the velocity of the impactor as a free parameter that is constrained from simple dynamical bounds.

From angular momentum conservation, we get the following relation between the impactor mass $m_i$ and its incident speed $v_i$ (Parisi & Brunini 1997), assuming that the impact is inelastic (Korycansky et al. 1990):

$$v_i = \frac{2m_N R_N^2}{5m_i b}\left(1+\frac{m_i}{m_N}\right)\sqrt{\Omega^2 + \frac{\Omega_0^2}{\left(1+\frac{m_i}{m_N}\right)^2\left(1+\frac{m_i}{3m_N}\right)^4} - \frac{2\Omega_0\Omega\cos\alpha}{\left(1+\frac{m_i}{m_N}\right)\left(1+\frac{m_i}{3m_N}\right)^2}}, \quad (1)$$

where $b$ is the impact parameter of the collision, $\mathbf{\Omega}$ the present spin angular velocity of Neptune, $\mathbf{\Omega}_0$ the spin angular velocity that Neptune would have today if the GC had not occurred, and $\alpha$ is the angle between $\mathbf{\Omega}$ and $\mathbf{\Omega}_0$. We take a minimum value of 1 $m_\oplus$ for the oligarchic mass to impact proto-Neptune and assume a maximum oligarchic mass of 4 $m_\oplus$ for comparison.

We get Neptune data from the JPL homepage. The current radius of Neptune $R_N$ is taken as 24,764 km and the mass of Neptune after the GC, $(m_i + m_N)$, is taken as its current mass of $102.44 \times 10^{24}$ kg. The spin period of Neptune is $T = 16.11$ h, thus $\Omega = 1.08338 \times 10^{-4}$ s$^{-1}$ ($\Omega = 2\pi/T$). Neptune's current obliquity is $29.58^o$. In the single stochastic impact approach $\alpha = 29.58^o$. The ice to rock ratio of Neptune is 85 %. Assuming that all the ice and rock are contained at the core of the planet at the time of the GC, the core mass of Neptune is $\sim 87.074 \times 10^{24}$ kg. We then compute the core radius of Neptune assuming that its core density $\rho_N$ and that of Uranus $\rho_U$ are the same at the end of their formation:

$$\frac{3}{4}\frac{m_C}{\pi R_C{}^3} = \rho_N = \rho_U = \frac{3}{4}\frac{m_{uc}}{\pi R_{UC}{}^3}, \qquad (2)$$

where $m_{uc} = 74.737 \times 10^{24}$ kg is the core mass of Uranus (Parisi & Brunini 1997) and $R_{UC} = 1.8 \times 10^4$ km its core radius (Bodeheimer & Pollak 1986). From Eq. (2), we get the core radius of Neptune, $R_C = 1.9 \times 10^4$ km. At the end of the formation of Uranus and Neptune, their gas envelopes extend until the accretion radius (e.g. Bodenheimer & Pollack 1986). Then, the radius of Neptune is $\sim 2.724 \times 10^7$ km (1,100 $R_N$), whereas its core containing most of the mass has a radius of only $1.9 \times 10^4$ km. In this situation, a collision onto the core is necessary for an inelastic collision to occur and to impart the required angular momentum (Korycansky et al. 1990). Since $b$ is an unknown quantity, we take its most probable value: $b = (2/3)R_C$ (Parisi et al. 2008).

To set constraints on the impactor speed $v_i$, we calculated the lower bound of $v_i$ ($v_{im}$) corresponding to a body within the Hill radius $R_H$ of Neptune and undergoing a free fall towards Neptune's core:

$$v_{im} = \sqrt{\frac{2Gm_C}{R_C} - \frac{2Gm_C}{R_H}}. \qquad (3)$$

The second term of Eq. (3) is negligible, and then, $v_{im}$ depends only slightly on $m_i$. Taking the average of $v_{im}$ for $m_i$ between 1-4 $m_\oplus$, we get $v_{im} \sim 24.758$ km s$^{-1}$.

The upper bound of $v_i$ ($v_{iM}$) for a body bound to the Solar System corresponds to a parabolic orbit with the impactor lying on the same orbital plane as proto-Neptune and moving in a direction opposite to Neptune's motion, including the acceleration caused by the planet (Parisi & Brunini 1997):

$$v_{iM} = \sqrt{\left[\sqrt{\frac{2GM_\odot}{a_N}} + \sqrt{\frac{GM_\odot}{a_N}}\right]^2 + \frac{2GM_N}{R_C}}, \qquad (4)$$

where $M_\odot$ is the mass of the Sun and $a_N$ the Neptunian orbital semiaxis at the time of the GC. Assuming that the GC with Neptune occurred at 30 AU, $v_{iM}$ is $\sim 30$ km s$^{-1}$. If Neptune had been formed between 12 and 30 AU (Dodson-Robinson et al. 2010, Benvenuto et al. 2009) and the GC had occurred before or during outward migration (Tsiganis et al. 2005), $v_{iM}$ would had been between 35 and 30 km s$^{-1}$. It should be noted that our estimate of $v_{iM}$ in Eq. (4) must be taken with care since big bodies velocity dispersion may increase so much that some fraction of them in the outer Solar System are ejected. Then, we take a maximum possible value for $v_{iM}$ of $\sim 40$ km s$^{-1}$.

We computed $v_i$ as a function of $T_0$ ($T_0 = 2\pi/\Omega_0$) through Eq. (1) for $m_i$ 1, 2.7, and 4 $m_\oplus$. Since $\alpha$ is a free parameter, we took six values of $\alpha$: $0^o$, $29.58^o$, $60^o$, $90^o$, $130^o$, and $170^o$. For $\alpha$ between $180^o$ and $360^o$, the results would be the same as for the interval $[0^o, 180^o]$ since Eq. (1) is an even function of $\alpha$. The cases with $\alpha \geq 60^o$ are less probable since it requires a very high initial obliquity before the GC. For each $\alpha$, $v_i(T_0)$ must fall between the bounds given by $v_{im}$ and $v_{iM}$. In Fig.1, we show the results for $m_i = 1$ $m_\oplus$. The permitted values of $T_0$ for $\alpha = 0^o$ (Fig.1a) are between 6.95 h and 8.80 h and $T_0 > 43.70$ h. If the present obliquity of Neptune was caused by a single stochastic impact (Fig.1b), the permitted values of $T_0$ are between 7.87 h and 11.14 h, and $T_0 > 34.52$ h. In Fig.1c, $T_0$ must be greater than 12.29 h. If $\alpha \geq 90^o$, $T_0$ must be greater than 30 h. The curves shift up and right as $\alpha$ increases. In Fig.2, we display the same as in Fig.1 for $m_i = 2.7$ $m_\oplus$. The intervals of permitted $T_0$ are [3.15, 4.49] h for $\alpha = 0^o$ (Fig.2a) and [6.28,15.66] h for $\alpha = 170^o$ (Fig.2f). For intermediate $\alpha$, the range of permitted $T_0$ falls between both cases. In Fig.3, we show the results for $m_i = 4$ $m_\oplus$, where we can see that the permitted $T_0$ are very low[1]. Then, we calculate the break-up speed $\Omega_{0b}$ given by

$$\frac{Gm_N}{R_{N0}^2} = \Omega_{0b}^2 R_{N0}, \qquad (5)$$

where $R_{N0} = R_N/(1 + m_i/3m_N)$ (Parisi & Brunini 1997).

From Eq. (5), we get the period $T_{ob}$ ($2\pi/\Omega_{0b}$) tabulated in Table 1. If $T_0 < T_{ob}$, the planet breaks up since centrifugal forces exceed gravitational forces. We then discard the cases shown in Fig.3a-c since the permitted $T_0$ are very close to $T_{ob} = 3.43$ h. It should be noted that the curves of $v_i$ in Figs. 1-3 shift to the left side as $m_i$ increases. Also, the cases for $\alpha \geq 60^o$ are unlikely, since a very high initial obliquity would be required. Then, for an impactor mass $m_i > 4$ $m_\oplus$ the permitted $T_0$ diminishes, while $T_{ob}$ increases (see Table 1). This implies that an impactor mass $m_i > 4$ $m_\oplus$ would refute the GC hypothesis since $T_0$ is less than $T_{ob}$ unless $\alpha$ is very large or $b \ll 2/3 R_C$. However, the probability that $b < R_C/3$ is 0.1 and may be discarded. For $b$ between $R_C/3$ and $2/3R_C$, the probability is 0.33, and it cannot be discarded. However, for these values of $b$, the curves of Fig.3 shift upwards by a factor between 1 and 2, and the permitted $T_0$ remain lower than five hours for Fig.3a-b. Thus, even for $b$ between $R_C/3$ and $2/3R_C$, the permitted $T_0$ are close to $T_{ob}$ unless $\alpha$ is large. Moreover, if the impactor is large, the most probable impact parameter is $b = 2/3(R_c + R)$ (R is the impactor radius). Then, the probability that b were less than $2/3\ R_c$ would be even lower.

---
[1] In all the cases, $v_i$ must be zero for $\alpha = 0^o$ and $T = T_0$. However, looking at the figures, the root of Eq. (1) shifts slightly to the left side of $T_0 = T$ as $m_i$ increases. In the development of Eq. (1), we considered an expression to first order in terms of the protoplanet radius increment for the mass increment $m_i/m_N$ (Parisi & Brunini 1997). For $m_i = 1$ $m_\oplus$, this approximation introduces an error of $\sim 10^{-3}$ in computing $m_i/m_N$, while for $m_i = 2.7$ $m_\oplus$, this error is $\sim 0.02$ and $\sim 0.05$ for $m_i = 4$ $m_\oplus$.

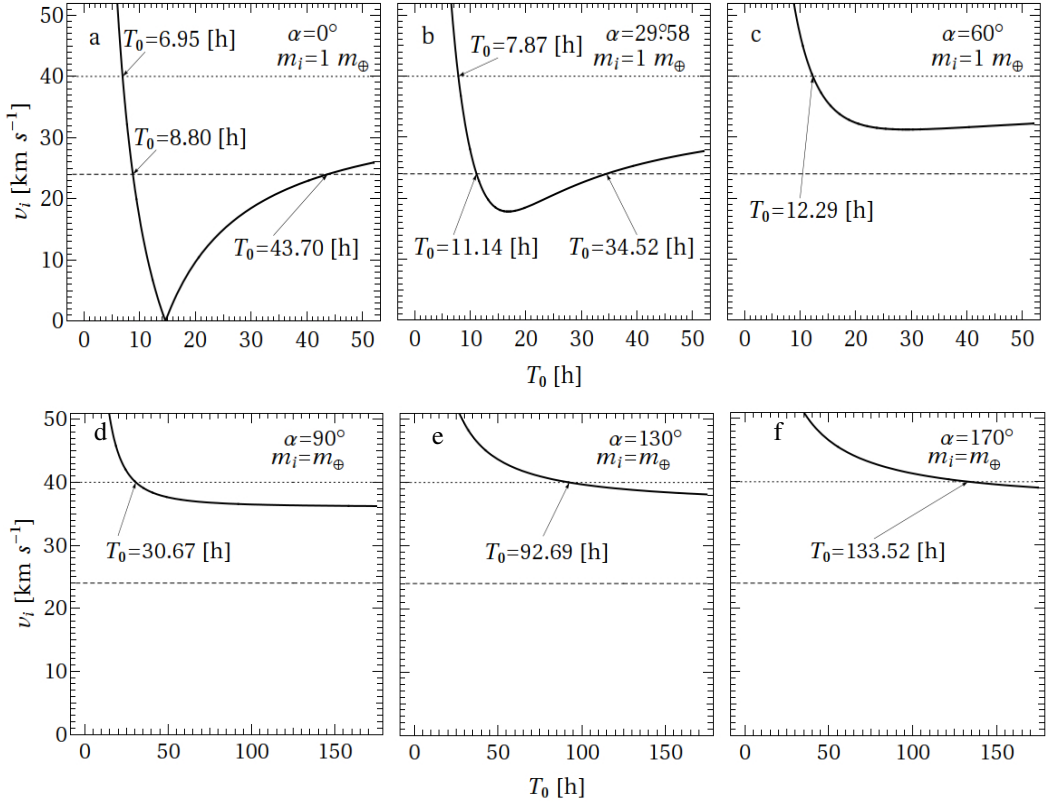

**Fig. 1.** Impactor incident speed $v_i$ as a function of the initial period $T_0$ for $m_i = 1\ m_\oplus$, obtained through angular momentum transfer for the most probable impact parameter $b = 2/3\ R_c$. a-$\alpha = 0°$, b-$\alpha = 29.58°$, c-$\alpha = 60°$, d-$\alpha = 90°$, e-$\alpha = 130°$, and f-$\alpha = 170°$. The upper constraint on $v_i$ ($v_{iM}$) is depicted by dotted lines while the lower constraint on $v_i$ ($v_{im}$) is shown by dashed lines. $v_{iM}$ and $v_{im}$ are obtained from simple dynamical constraints.

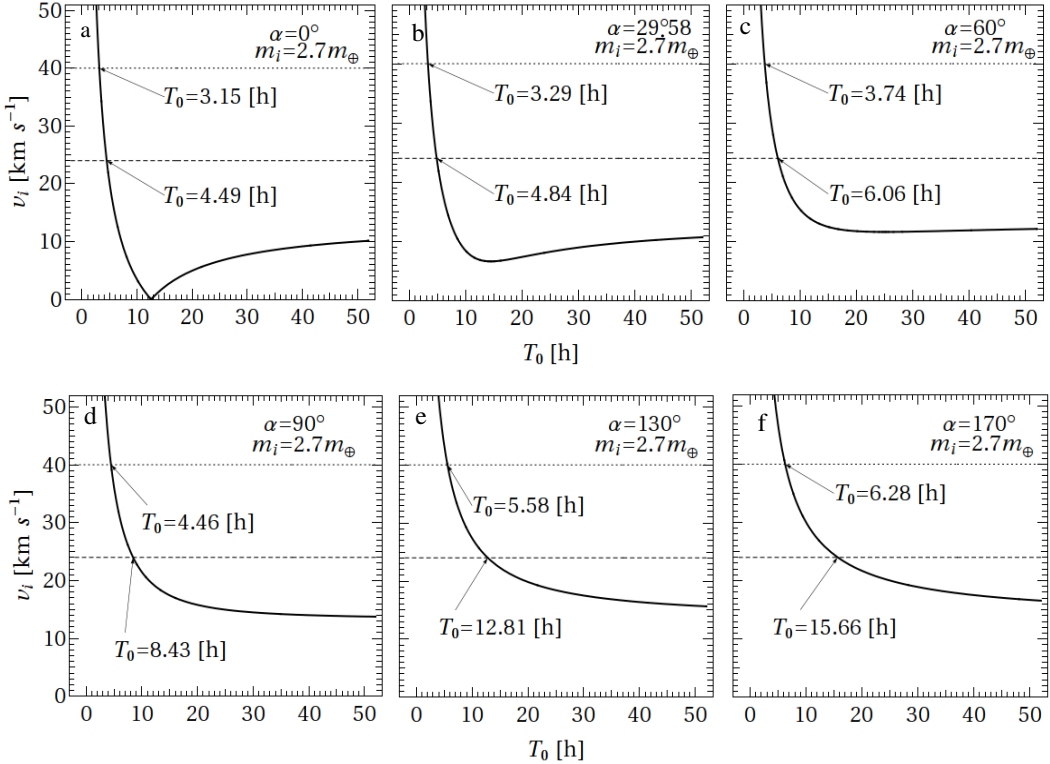

**Fig. 2.** The same as in Fig.1 for $m_i = 2.7\ m_\oplus$.

**Table 1.** Break-up period of proto-Neptune before the GC for different impactor masses.

| $m_i$ [$m_\oplus$] | 1 | 2 | 3 | 4 | 5 | 6 | 7 | 8 |
|---|---|---|---|---|---|---|---|---|
| $T_{ob}$ [h] | 2.76 | 2.95 | 3.17 | 3.43 | 3.75 | 4.13 | 4.61 | 5.23 |

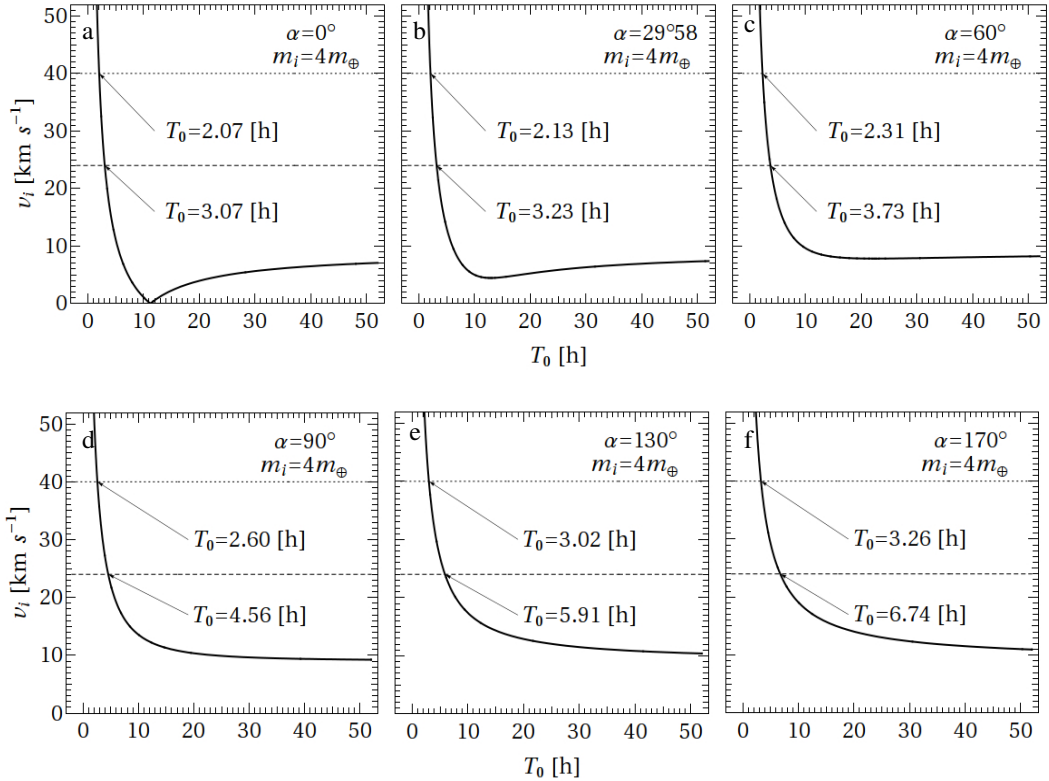

**Fig. 3.** The same as in Fig.1 for $m_i = 4\ m_\oplus$.

We may then conclude that an impact with a mass higher than 4 $m_\oplus$ probably could not have occurred since it would be improbable that it could reproduce the present rotational properties of Neptune. It might be understood as an upper bound for the oligarchic masses in the trans-Saturnian region at the end of Neptune formation.

## 3. Irregular satellites: May they set constraints on giant impacts?

### 3.1. Their origin: how and when irregular satellites of giant planets were captured remains open

Rich systems of irregular satellites of the giant planets have been discovered very recently. The study of their origin is very important because it can provide constraints on the formation process of giant planets and may probe the properties of the primordial planetesimal disk (Parisi et al. 2008, Vokrouhlicky et al. 2008). Neptune has at present seven irregular satellites: I Triton, II Nereid, IX Halimede, XI Sao, XII Laomedeia, XIII Neso, and X Psamathe (Holman et al. 2004, Sheppard et al. 2006b). Some of their physical and orbital parameters are shown in Table 2.

Irregular satellites of giant planets are characterized by eccentric (highly tilted with respect to the parent planet equatorial plane) and in some case retrograde, orbits. These objects cannot have formed by circumplanetary accretion as the regular satellites but they are likely products of an early capture of primordial objects from heliocentric orbits, probably in association with planet formation itself (Jewitt & Sheppard 2005). It is possible for an object circling the sun to be temporarily trapped by a planet. In terms of the classical three-body problem, this type of capture can occur when the object passes through the interior Lagrangian point, $L_2$, with a very low relative velocity. But, without any other mechanism, such a capture is not permanent and the objects will eventually return to a solar orbit after several or several hundred orbital periods. To turn a temporary capture into a permanent one requires a source of orbital energy dissipation and needs for particles to remain inside the Hill sphere long enough for the capture to be effective. Although giant planets currently have no efficient mechanism of energy dissipation for permanent capture, several mechanisms may have operated at their formation epoch: 1) *gas drag* in the solar nebula or in an extended, primordial planetary atmosphere or in a circumplanetary disk (Pollack et al.1979, Cuk & Burns 2003). 2) *Pull-down capture* caused by the mass growth and/or orbital expansion of the planet, which expands its Hill sphere (Brunini 1995, Heppenheimer & Porco 1977). 3) *Collisional interaction* between two planetesimals passing near the planet or between a planetesimal and a regular satellite. This mechanism, the so-called *break-up* process, leads to the formation of dynamical groupings (e.g. Colombo & Franklin 1971, Nesvorny et al. 2004). After a break-up the resulting fragments of each progenitor form a population of irregular satellites with similar surface composition, i.e. similar colors and irregular shapes, i.e. light curves of wide amplitude (Maris et al. 2001, Maris et al. 2007). 4) *Collisionless interactions* between a massive planetary satellite and guest bodies (Tsui 1999) or between the planet and a binary object (Agnor & Hamilton 2006, Vokrouhlicky et al. 2008). 5) *Scattering by an eccentric Triton* applicable to the Neptunian system (Goldreich et al.1989, Cuk & Gladman 2005), and 6) *collisionless interactions within the framework of the Nice model*, where capture occurs during migration through three-body interactions during close encounters between the giant planets (Nesvorny et al. 2007, Bottke et al. 2010). There might have been several encounters and stochastic collisions among protoplanets during migration, all of which could have both removed and captured irregular satellites. It was shown that the last captured population of irregular satellites by three-body encounters in the framework of the Nice model were very likely significantly larger than what we observe today. These populations had then experienced rapid collisional evolution and almost self destructed, evolving into quasi-steady state size frequency distributions with extremely shallow power-law slopes for D> 10 km (Bottke et al. 2010).

The knowledge of colors and of the size and shape distribution of irregular satellites is important for knowing their relation to the precursor population. It brings valuable clues to investigate if they are collisional fragments from break-up processes occurring at the planetesimal disk and thus has nothing to do with how they were individually captured later by the planet, or if they are collisional fragments produced during or after the capture event (Nesvorny et al. 2003, 2007, Bottke et al. 2010). Despite many studies in past years, inconsistencies are present between colors derived by different authors for the irregular satellites of Uranus and Neptune (Maris et al. 2001, Maris et al. 2007, Grav et al. 2004). Moreover, if the precursors of the Uranian and Neptunian irregular satellites are objects of the Kuiper Belt or C-,P-, and D-type asteroids is still a matter of debate (Bottke et al. 2010, Levison et al. 2009). Even the recent proposed capture models run into trouble using the size frequency distribution of irregular satellites as constraints (Nesvorny et al. 2007, Bottke et al. 2010). An intensive search for fainter irregular satellites and a long term program of observations to recover light curves, colors, and phase-effect information in a self-consistent manner is needed. Constraints on the different dissipative mechanisms for the permanent capture of irregular satellites should be investigated in the framework of the different models proposed for the formation of the ice giants.

Although, recent models study the capture of irregular satellites by close encounters between giant planets in the framework of the Nice model, there is very little work on the capture and loss of irregular satellites by stochastic collisions (Parisi et al. 2008). Here, we investigate whether the present population of Neptunian irregular satellites could have been captured by stochastic impacts and/or if they could have survived to stochastic impacts. We attempt to set constraints on the occurrence of giant impacts at the end of the formation of the ice giants, on the impactor masses, and on the epoch of the capture/origin of the Neptunian irregular satellites.

### 3.2. The Neptunian irregular satellites were captured before, after, or during stochastic impacts?

All stochastic impacts add angular momentum to the random component of planetary spin and transfer impulse to the planet and its satellite system. Although the transfer of impulse at any collision onto the planet affects the whole system, the transfer of angular momentum affects only the rotational properties of the planet but not the orbital or rotational properties of its satellites.

The impulse accumulated by successive collisions with a distribution of small or/and large planetesimals requires ad-hoc assumptions about the unknown planetesimal mass distribution, the unknown initial surface mass density, and the unknown regime of growth. In the GC scenario, we avoid the necessity of quantifying these parameters, since we assume that the orbital properties of the irregular satellites of Neptune before the GC are unknown as are the physical and dynamical properties of oligarchs. These initial free parameters are constraint by the model from the current orbital properties of the irregular satellites and using simple general dynamical bounds.

The transfers of six of the seven irregular satellites of Neptune to their current orbits by the GC are computed following a somewhat more improved procedure than in Parisi et al. (2008). Irregular satellites may be pushed to outer or inner orbits. Moreover, they may be captured or unbound by the GC. We attempt to set constraints on their initial orbital properties, the impactor mass, and the epoch of their origin. We do not include Triton in our treatment, since this satellite is assumed to have a different origin than the other irregular satellites of Neptune owing to its high mass, small orbital semiaxis and zero orbital eccentricity (see Table 2).

Just before the GC, the square of the orbital velocity $v_1$ of a preexisting satellite of negligible mass is given by

$$v_1^2 = Gm_N\left(\frac{2}{r} - \frac{1}{a_1}\right), \qquad (6)$$

where $r$ is the position of the satellite on its orbit at the moment of the GC and $a_1$ its orbital semiaxis.

After the GC, the satellite is transferred to another orbit with semiaxis $a_2$ acquiring the following square of the velocity:

$$v_2^2 = G(m_N + m_i)\left(\frac{2}{r} - \frac{1}{a_2}\right). \qquad (7)$$

We set $v_1^2 = A\ v_e^2$ and $v_2^2 = B\ (1 + m_i/m_N)\ v_e^2$, where $A$ and $B$ are arbitrary coefficients ($0 < A \le 1$, $B > 0$), $v_e$ being the escape velocity at $r$ before the GC.

**Table 2.** Orbital and physical parameters of the Neptunian irregulars taken from JPL (http://ssd.jpl.nasa.gov).

| Irregular Satellites | $r_s^*$ | $\rho_s^{**}$ | Magnitude | Albedo | $a^x$ | e | $i^+$ | $P^{++}$ |
|---|---|---|---|---|---|---|---|---|
| I    Triton    | 1,352.6 | 2.06 | 13.5 $V_o$ | 0.76 | 14.33   | 0.0000 | 156.885 | 5.88 |
| II   Nereid    | 170.0   | 1.50 | 19.7 $V_o$ | 0.16 | 222.65  | 0.7507 | 7.230   | 360.13 |
| IX   Halimede  | 31.0    | 1.50 | 19.2 R     | 0.04 | 670.77  | 0.2646 | 112.712 | 1879.08 |
| XI   Sao       | 22.0    | 1.50 | 24.5 R     | 0.04 | 897.60  | 0.1365 | 53.483  | 2912.72 |
| XII  Laomedeia | 21.0    | 1.50 | 25.5 R     | 0.04 | 951.66  | 0.3969 | 37.874  | 3171.33 |
| XIII Neso      | 30.0    | 1.50 | 24.6 R     | 0.04 | 1990.19 | 0.5714 | 136.439 | 9740.73 |
| X    Psamathe  | 20.0    | 1.50 | 25.5 R     | 0.04 | 1942.17 | 0.3809 | 126.312 | 9074.30 |

\* The mean satellite radius $r_s$ is given in km.
\*\* The mean satellite density $\rho_s$ is given in g cm$^{-3}$.
x The orbital semiaxis a is in units of $R_N$= 24,764 km.
+ The orbital inclination i is measured with respect to the ecliptic.
++ The orbital period P is given in days.

The semiaxis of the satellite orbit before ($a_1$) and after ($a_2$) the GC verify the following simple relations:

$$a_1 = \frac{r}{2(1-A)} \quad , \quad a_2 = \frac{r}{2(1-B)} . \qquad (8)$$

If $A < B$ then $a_1 < a_2$. In the special case of $B = 1$, the orbits are unbound from the system. If $A > B$ then $a_1 > a_2$, the initial orbit is transferred to an inner orbit, providing a mechanism for the permanent capture of the irregular satellite even from an unbound orbit ($A \geq 1$). When $A = B$, the orbital semiaxis remains unchanged ($a_1 = a_2$).

The position $r$ of the satellite on its orbit at the epoch of the impact may be expressed in the following form (Bunini et al. 2002, Parisi et al. 2008):

$$r = \frac{2\,G\,m_N}{(\Delta V)^2} \left[ \frac{B'-A}{\sqrt{A}\cos\Psi \pm \sqrt{(B'-A)+A\cos^2\Psi}} \right]^2, \qquad (9)$$

with $B' = B(1 + m_i/m_N)$, and $\Psi$ is the angle between $\mathbf{v_1}$ and the orbital velocity change $\Delta V$ imparted to Neptune by the GC. The value of $\Delta V$ is obtained through impulse conservation considerations at collision (Parisi & Brunini 1997):

$$(m_i + m_N)\Delta V = m_i \mathbf{v_i}. \qquad (10)$$

Upper bounds on $a_1$ ($a_{1M}$) and $a_2$ ($a_{2M}$) are obtained from Eqs. (8) and (9) with $\Psi=180^o$, i.e., assuming the impact in the direction opposite to the orbital motion of the satellites and taking the positive sign of the square root:

$$a_{1M} = \frac{G\,m_N\,(B'-A)^2}{(\Delta V)^2\,(1-A)(\sqrt{B'}-\sqrt{A})^2}$$
$$a_{2M} = \frac{G\,m_N\,(B'-A)^2}{(\Delta V)^2\,(1-B)(\sqrt{B'}-\sqrt{A})^2} . \qquad (11)$$

The minimum eccentricity of the orbits before the collision is given by

$$e_{1min} = 2(1-A) - 1 \quad if \quad A \leq 0.5$$
$$e_{1min} = 1 - 2(1-A) \quad if \quad A > 0.5, \qquad (12)$$

while the minimum eccentricity of the orbits after the collision is

$$e_{2min} = 2(1-B) - 1 \quad if \quad B \leq 0.5$$
$$e_{2min} = 1 - 2(1-B) \quad if \quad B > 0.5. \qquad (13)$$

Using the current orbital elements of the Neptunian irregular satellites as initial conditions (see Table 2), their orbital evolution over a period of $10^5$ yrs is computed using the integrator evorb 13 of Fernández et al. (2002), where the perturbations of the Sun, Jupiter, Saturn and Uranus are included. The minimum, mean, and maximum eccentricities and semiaxes are shown in Table 3 for all the Neptunian irregular satellites with the exception of Triton. We assume that gravitational perturbations are the only effect capable of altering the orbital elements of the irregular satellites during Solar System evolution after the GC since gas drag caused by an ice giant's envelope proves negligible (Parisi et al. 2008).

We compute $a_{1M}$ and $a_{2M}$ from the upper and lower bounds of $\Delta V$ obtained through Eq. (10) using $v_i=v_{iM}$ and $v_i=v_{im}$, respectively. We take $v_{iM}$= 40 km s$^{-1}$ and $v_{im}$ from Eq. (3). We take $m_i$= 1 $m_\oplus$ as the minimum impactor mass. For each $A$, we then calculate the values of $B$ ($B = B'/(1 + m_i/m_N)$) corresponding to the transfers to $a_{2M}= [a_{max}, a_{min}]$, where $a_{max}$ and $a_{min}$ are the current maximum and minimum semiaxes of the irregular satellites tabulated in Table 3. Introducing these values of $B$ in Eq. (13), we get the minimum possible values of $e_{2min}$, $e_{2m}$, that the orbit of each irregular satellite may acquire at impact for each initial condition $A$.

The transfer of each irregular satellite from its original orbit to the present one is only possible for those values of $(A, B)$ which satisfy the condition $e_{2m} < e_{max}$ ($e_{max}$ tabulated in Table 3). Transfers with $A < B$, imply that the irregular satellite belongs from an inner orbit, while transfers with $A > B$ imply that the irregular satellite belongs from an outer orbit; i.e., the GC may produce the permanent capture of a body undergoing a temporary capture or even in heliocentric orbit ($A \geq 1$).

The transfers of Halimede and Nereid are shown in Fig.4, those of Sao and Laomedeia in Fig.5, and the transfers of Neso and Psamathe are displayed in Fig.6. The lower curve in Figs.4-

**Table 3.** Variation in the orbital eccentricity and semiaxis (in units of $R_N = 24{,}764$ km) of the Neptunian irregulars due to Solar and giant planet perturbations over a period of $10^5$ yrs. $\Delta_{a_{min}} = a_{mean} - a_{min}$ and $\Delta_{a_{max}} = a_{max} - a_{mean}$.

| Satellite | $e_{min}$ | $e_{mean}$ | $e_{max}$ | $a_{min}$ | $a_{mean}$ | $a_{max}$ | $\Delta^*_{a_{min}}$ | $\Delta^{**}_{a_{max}}$ |
|---|---|---|---|---|---|---|---|---|
| Nereid | 0.7426 | 0.7496 | 0.7566 | 222.633 | 222.659 | 222.687 | 0.026 | 0.028 |
| Halimede | 0.2073 | 0.4331 | 0.9026 | 668.865 | 670.297 | 672.025 | 1.432 | 1.728 |
| Sao | 0.0628 | 0.2839 | 0.6619 | 892.175 | 896.893 | 902.070 | 4.718 | 5.177 |
| Laomedeia | 0.2648 | 0.4015 | 0.5482 | 942.266 | 948.295 | 955.526 | 6.029 | 7.231 |
| Neso | 0.0942 | 0.4140 | 0.8803 | 1,919.790 | 2,013.260 | 2,130.300 | 93.470 | 117.040 |
| Psamathe | 0.0482 | 0.3970 | 0.9144 | 1,832.740 | 1,912.260 | 2,015.540 | 79.520 | 103.280 |

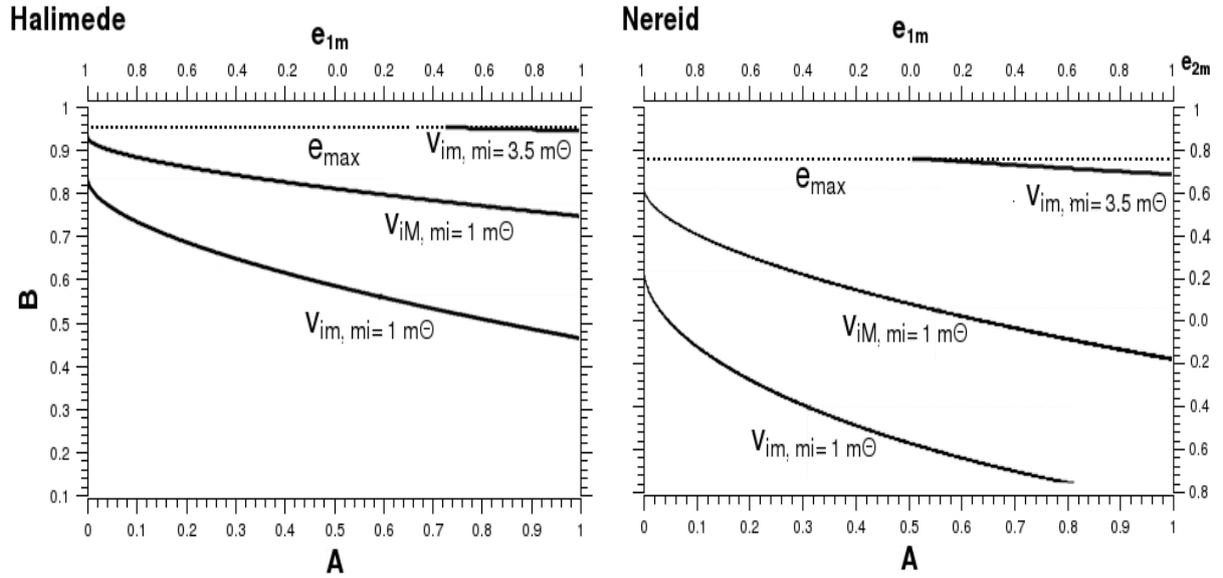

**Fig. 4.** The transfers capable of producing the present orbits of Halimede (left panel) and Nereid (right panel). Lower curve corresponds to the minimum impactor speed $v_{im}$ and an impactor mass $m_i = 1\ m_\oplus$. Intermediate curve corresponds to the maximum impactor speed $v_{iM}$ and $m_i = 1\ m_\oplus$. Upper curve corresponds to $v_{im}$ and $m_i = 3.5\ m_\oplus$. The maximum current eccentricity $e_{max}$ is shown by a dotted line. For Halimede the upper curve and $e_{max}$ overlap. A (B) is the square of the ratio of the satellite's speed just before (after) the impact on the escape velocity at the satellite's location just before (after) the impact. $e_{1m}$ ($e_{2m}$) is the minimum eccentricity of the orbits before (after) collision.

6 represents the transfers for $v_i = v_{im}$ and $m_i = 1\ m_\oplus$. It should be noted that by minimizing $v_i$ and $m_i$, we are maximizing the number of transfers. The real eccentricity that the irregular satellite may acquire for each $v_i$ and $m_i$ may take any value between $e_{2m}$ and $e_{max}$. The probability of the existence of the irregular satellite prior to the GC diminishes as $e_{2m}$ approaches $e_{max}$.

In Fig.4, the intermediate curve represents the transfers for $m_i = 1\ m_\oplus$ and $v_i = v_{iM}$. The transfers for $m_i = 1\ m_\oplus$ and $v_{im} \leq v_i \leq v_{iM}$ fall between the lower and intermediate curves. The upper curve corresponds to $v_{im}$ and $m_i = 3.5\ m_\oplus$, and $e_{max}$ and the upper curve overlap. Halimede and Nereid then set an upper bound on $m_i$ of 3.5 $m_\oplus$. Within this mass limit, the GC itself could provide a mechanism for the capture of Halimede and Nereid.

In Fig.5, the transfers for Sao are shown in the left panel. There are no transfers for $v_{iM}$ for any impactor mass. The upper curve shows the transfers for $v_{im}$ and $m_i = 1.6\ m_\oplus$, which is very close to $e_{max}$. There is no transfer for $m_i \geq 1.7\ m_\oplus$. Sao set an upper bound of 1.7 $m_\oplus$ on the impactor mass. The transfers of Laomedeia are displayed in the right panel. The upper curve displays the transfers for $v_{im}$ and $m_i = 1.35\ m_\oplus$, which are the same as for $v_{iM}$ and $m_i = 1\ m_\oplus$. Laomedeia set an upper bound on $m_i$ of 1.4 $m_\oplus$. Within this mass limit, the GC itself could provide a mechanism for the capture of Sao and Laomedeia.

The transfers of Psamathe are shown in the left panel and those of Neso in the right panel of Fig.6. The intermediate curve corresponds to the transfers for $v_{iM}$ and $m_i = 1\ m_\oplus$. The upper curve corresponds to $v_{im}$ and $m_i = 2.3\ m_\oplus$ (Psamathe), and $m_i = 1.8\ m_\oplus$ (Neso). Psamathe set an upper bound on $m_i$ of 2.3 $m_\oplus$, while Neso of 1.8 $m_\oplus$. Within this mass limit, the GC itself could provide a mechanism for the capture of Psamathe and Neso. The transfers of Fig.6 have been computed over a range of 200 $R_N$ around the current orbital semiaxes of both irregular satellites (see Table 3). However, $e_{2m}$ lies close to $e_{max}$ for $v_i > v_{im}$ and $m_i > 1\ m_\oplus$. This result makes the existence of Neso and Psamathe prior to the GC, as well as their capture by the GC, a possible but not very probable event.

Parisi et al. (2008) carried out a somewhat similar model of the Uranian system. They obtained strong constraints for two

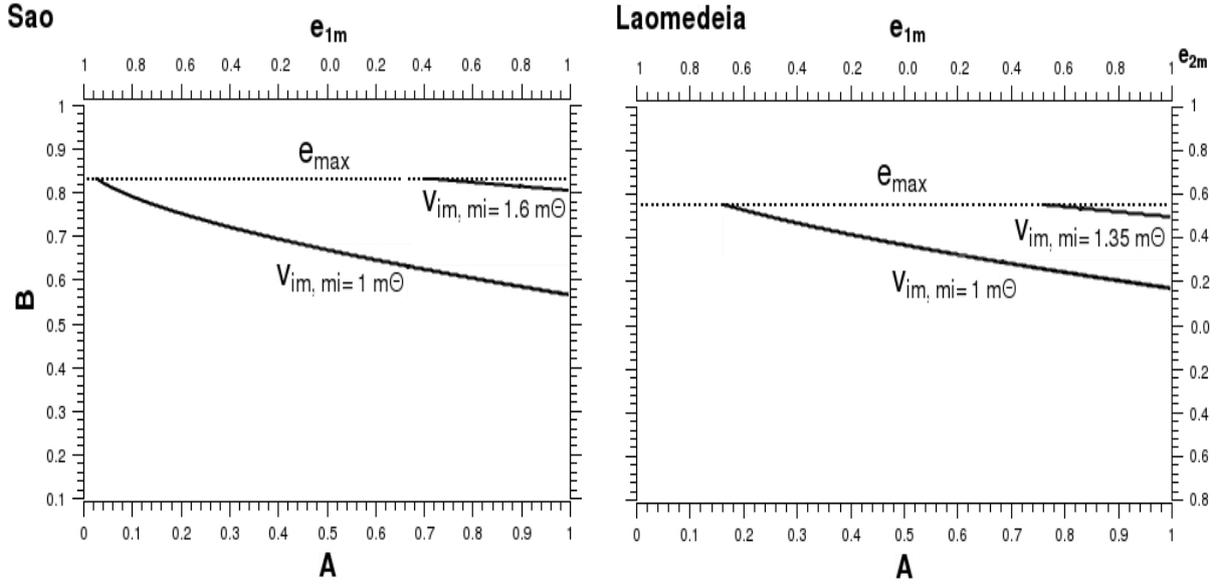

**Fig. 5.** The transfers capable of producing the present orbits of Sao (left panel) and Laomedeia (right panel). Lower curve corresponds to the minimum impactor speed $v_{im}$ and an impactor mass $m_i = 1\ m_\oplus$. Upper curve left panel corresponds to $v_{im}$ and $m_i = 1.6\ m_\oplus$ which is the same as for $v_{iM}$ and $m_i = 1\ m_\oplus$. Upper curve right panel corresponds to $v_{im}$ and $m_i = 1.35\ m_\oplus$. There is no transfer for $v_{iM}$ and $m_i = 1\ m_\oplus$. The maximum current eccentricity $e_{max}$ is shown by a dotted line. A (B) is the square of the ratio of the satellite's speed just before (after) the impact on the escape velocity at the satellite's location just before (after) the impact. $e_{1m}$ ($e_{2m}$) is the minimum eccentricity of the orbits before (after) collision.

Uranian irregular satellites: Prospero and Trinculo. The only transfers for both irregular satellites were close to the pericenter of an eccentric initial orbit ($e_{1m} > 0.58$ for Prospero and $e_{1m} > 0.62$ for Trinculo). $e_{2m}$ for Trinculo resulted in the range [0.16-0.23], very close to $e_{max}$(0.237). Moreover, for Prospero $e_{2m}$ was $\sim e_{max}$ (0.571). This result implied that Prospero could not survive stochastic impacts. Either Prospero was created after the epoch of giant impacts or giant impacts did not occur.

In this paper, Laomedeia set an upper constraint on the impactor mass of 1.4 $m_\oplus$. Assuming that the Neptunian irregular satellites belong to a common origin, then they set an upper constraint of 1.4 $m_\oplus$ on the impactor masses in the trans-Saturnian region at the end of Neptune formation.

Assuming that giant impacts occurred beyond oligarchy, our results imply that either the oligarchic masses in the trasn-Saturnian region at the end of Neptune formation were less that 1.4 $m_\oplus$, or the current Neptunian irregular satellites had to be captured after the end of stochastic impacts. Alternatively, parent objects with orbits different to the irregular satellites' current orbits could have been captured by any process before or during stochastic impacts without restrictions on the impactor mass, but the present population of irregular satellites should then be the result of their later collisional evolution. This possibility would agree with the results of Bottke et al. (2010).

## 4. Discussion

It has long been known that dynamical times in the trans-Saturnian MMSN are so long that core growth takes more than 15 Myr. Observations of young, solar-type stars suggest that circumstellar disks dissipate on a timescale of few Myr (e.g. Briceño et al. 2001). Different models and scenarios have been proposed to account for the formation of the ice giants within the protoplanetary disk lifetime. Modern models shorten the timescale for giant planet formation if taking a higher initial surface density into account well above that of the MMSN and/or the formation of all giant planets in an inner compact configuration (Dodson-Robinson et al. 2010, Benvenuto et al. 2009, Thommes et al. 2003, Tsiganis et al.2005).

Core accretion models of giant planet formation in situ (Pollack et al.1996) and core accretion models with the giant planets forming at the initial semiaxes of the Nice model (Dodson-Robinson et al.2010, Benvenuto et al. 2009) compute the solid component because of the accretion of planetesimals. However, no core accretion model has at present, considered the angular momentum transfer during accretion and the possible accretion of other protoplanets onto the core with the exception of Korycansky et al. (1990) who carried out a hydrodynamical model of a giant impact on Uranus.

Modern scenarios have several difficulties to overcome, and inconsistencies among the different models and scenarios are still present. We discuss four important issues below.

1) Most core accretion models (e.g. Dodson-Robinson et al. 2010), assume planetesimals encountering the core at the Hill velocity; i.e., they do not take the stirring of these bodies into account. The Hill velocity $v_H$ is a characteristic velocity associated with the Hill radius $R_H$ and is defined as $v_H = \Omega R_H$, where $\Omega$ is the orbital frequency around the Sun. A speed of the surrounding planetesimals higher than $v_H$ would diminish the rate of growth of the ice giants in the massive planetesimals disk: either they had formed in situ or at 10-20 AU.

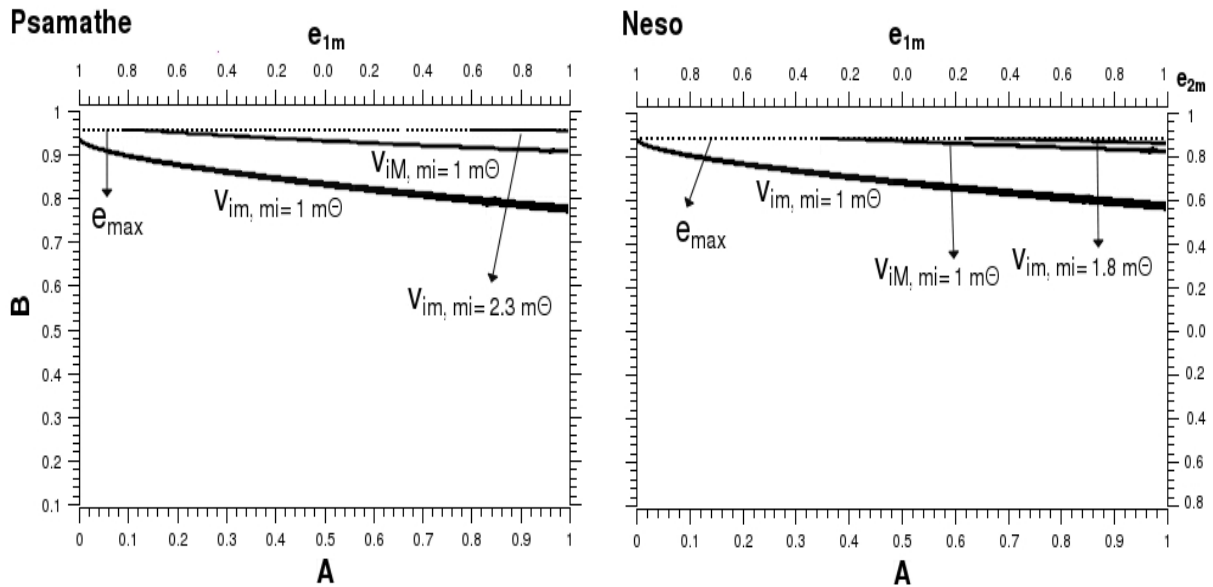

**Fig. 6.** The transfers capable of producing the present orbits of Psamathe (left panel) and Neso (right panel). Lower curve corresponds to the minimum impactor speed $v_{im}$ and an impactor mass $m_i = 1$ $m_\oplus$. Intermediate curve corresponds to the maximum impactor speed $v_{iM}$ and $m_i = 1$ $m_\oplus$. Upper curve left panel corresponds to $v_{im}$ and $m_i = 2.3$ $m_\oplus$. Upper curve right panel corresponds to $v_{im}$ and $m_i = 1.8$ $m_\oplus$. The maximum current eccentricity $e_{max}$ is shown by a dotted line. In both panels the upper curve and $e_{max}$ overlap. A (B) is the square of the ratio of the satellite's speed just before (after) the impact on the escape velocity at the satellite's location just before (after) the impact. $e_{1m}$ ($e_{2m}$) is the minimum eccentricity of the orbits before (after) collision.

2) An initial mass 5-10 the MMSN is required to shorten the timescale of ice giants formation, but within a 10 MMSN, Jupiter falls like a stone into the Sun due to type III migration (Crida 2009).

3) If oligarchic masses can reach the isolation mass in 1-10 Myrs at 10-30 AU is unknown. The final mass of the oligarchs in the outer solar System remains an open question. Without collisions among oligarchs, the mass of the core of proto-Uranus and proto-Neptune in the frame of the MMSN is the mass of an oligarch; i.e., much low than the present solid cores of Uranus and Neptune (either the ice planets had formed in situ or between 10-20 AU). Collisions among oligarchs takes a very long time to form the ice giants cores. Then, an initial surface density ~ 5-10 times that of the MMSN would be required to produce oligarchs with masses similar to those of the ice giants cores. However, a solid surface density of this size would lead to the formation of about five ice giants instead of two, which occurred with the other three giants; i.e., whether they were ejected or if they simply were spread out being all retained is a matter of debate (Goldreich et al. 2004, Dodson-Robinson et al. 2010, Ford & Chiang 2007, Levison & Morbidelli 2007). In the last case, then, where are they? However, Thommes et al. (2003) have shown that even in a disk 10 times the MMSN, oligarchs do not have time to reach their isolation mass in the outer Solar System, and even an Earth mass at the orbit of Uranus by 10 Myrs is implausible.

4) Actually, the mass of the planetesimals from which Uranus and Neptune accreted is a matter of debate. Goldreicih et al. (2004) suggest that, in the dispersion dominate regime the velocity dispersion of small bodies $u$ could exceed their surface escape velocity, and thus collisions at these speeds would be destructive. They speculate that before the oligarchs become planets, small bodies could fragment, reducing their size. Smaller size would imply smaller $u$, hence faster accretion, thus returning to the case $u \leq v_H$ at isolation. However, there is no reason planetesimals should fragment in the dynamically cold disk required to produce Uranus and Neptune in situ (Dodson-Robinson et al. 2010). It is usually believed that the planetesimal distribution follows a power law of the type $dn \propto m^{-\alpha} dm$ (e.g. Benvenuto et al. 2009 and references therein). Indeed, Jupiter family comets follow a distribution of this type with $\alpha = 11/6$. Benvenuto et al. (2009) computed the formation of Uranus and Neptune assuming the initial semiaxes distribution of the Nice model taking $\alpha = 2.5$. They approached the continuous distribution to a discrete distribution with planetesimal sizes in the range 30 m 100 km. However, Dodson-Robinson et al. (2010) computed the accretion of Uranus and Neptune in the frame of the Nice model, too, but considered planetesimals with radii of several hundred kilometers since they assumed the theory of planetesimal formation based on the streaming instability which produces planetesimals of about 600 km (Johansen et al. 2007). Although the streaming instability has several benefits, observational evidence (e.g., comet outbursts and chondrules) shows a fractal structure of the planetesimals consistent with coagulation processes for planetesimal formation (Paszun & Dominik 2009). According to Morbidelli et al. (2009), asteroids were born big, suggesting that the minimal size of the planetesimals was ~ 100 km.

In summary, it is not known whether the ice planets formed in situ, or well inside the 20 UA and/or if the initial mass of the nebula was that of the MMSN or much larger. Moreover, the mass of the planetesimals from which Uranus and Neptune accreted remains a matter of debate. It is necessary to look for independent ways of setting constraints on models of ice giant planet formation.

Our models are independent of unknown parameters, such as the mass and distribution of the planetesimals, the location at which Uranus and Neptune were formed, the Solar Nebula initial surface mass density, and the regime of growth. Our constraints on the oligarchic masses may be used to set constraints on planetary formation scenarios.

## 5. Conclusions

We have modeled the angular momentum transfer to proto-Neptune and the impulse transfer to its irregular satellites by the last stochastic collision (GC) between the protoplanet and an oligarchic mass. Since stochastic impacts are aleatory and their number is uncertain, the rotational properties of proto-Neptune and the orbital properties of the Neptunian system before the GC are taken as initial free parameters in our model and are constrained using the present rotational parameters of Neptune and the present orbital and physical properties of Neptune and its irregular satellite population.

Assuming a minimum impactor mass $m_i \sim 1\, m_\oplus$, the mass of the last impactor $m_i \leq 4\, m_\oplus$ is required to account for the current spin properties of Neptune. This result is invariant: Neptune had formed either in situ or between 10-20 AU and does not depend on the possible occurrence of the stochastic impact during or after migration. A collision with $m_i > 4\, m_\oplus$ could not have occurred since it cannot reproduce the present rotational and physical properties of the planet, unless the impact parameter of the collision was very small. The formation of Neptune as the result, for instance, of collisional accretion between two similar oligarchs with masses $\sim 7\, m_\oplus$ seems to be unlikely. Assuming the occurrence of giant impacts as the cause of Neptune's obliquity, the $4\, m_\oplus$ mass limit must be understood as an upper bound for the oligarchic masses in the trans-Saturnian region at the end of Neptune's formation.

The study of the origin of irregular satellites of giant planets is very important because it puts constraints on formation processes of giant planets and may probe the properties of the primordial planetesimal disk from which irregular satellites were captured. Although recent models study the capture of irregular satellites by close encounters between giant planets in the framework of the Nice model (Bottke et al. 2010), there is very little work on their capture and loss by stochastic collisions (Parisi et al. 2008).

We found that either the mass of the last impactor on Neptune was less than $\sim 1.4\, m_\oplus$, or the present Neptunian irregular satellites had to be formed or captured after the end of stochastic impacts. This result is invariant: Neptune had formed either in situ or between 10-20 AU and does not depend on the possible occurrence of the stochastic impacts during or after migration.

If the present population of irregular satellites were captured after the end of stochastic impacts, the mechanism of capture able to operate at such late stages in the scenario of the formation of the planet should be investigated. Alternatively, parent objects with orbits different from the irregular satellites' current orbits could have been captured by any process before or during stochastic impacts, but the present population of irregular satellites should be the result of their later collisional evolution. This possibility would agree with the results of Bottke et al. (2010).

Colors are an important diagnostic tool in attempting to unveil the physical status and the origin of the Neptunian irregular satellites. In particular it is very important to assess whether it is possible to define subclasses of irregular satellites just by looking at colors and by comparing colors of these bodies with colors of minor bodies in the outer Solar System. The knowledge of the primordial population of irregular satellites could offer valuable clues to know whether Neptune was formed in situ or between 10-20 AU. An intensive search for fainter irregular satellites and a long term program of observations able to recover lightcurves, colors, and phase effects in a "self consistent" manner is mandatory.

*Acknowledgements.* MGP research was supported by Intituto Argentino de Radioastronomía, IAR-CONICET, and by CONICET grant PIP 112-200901-00461, Argentina. LdV research was supported by CONICYT Chile and DAS Universidad de Chile.